\documentclass[journal]{IEEEtran}

\usepackage{amsmath,amssymb,eucal,graphicx}
\usepackage{epsfig}
\usepackage{exscale}
\usepackage{verbatim}
\usepackage{amsmath}
\usepackage{amsfonts}
\usepackage{amssymb}
\usepackage{graphicx}
\usepackage{color}
\usepackage{setspace}
\usepackage{balance}

\newfont{\bbb}{msbm10 scaled 500}

\newfont{\bb}{msbm10 scaled 1100}

\newcommand{\EE}{\mbox{\bb E}}

\newcommand{\bv}{{\bf b}}

\newcommand{\hv}{{\bf h}}

\newcommand{\nv}{{\bf n}}

\newcommand{\sv}{{\bf s}}

\newcommand{\vv}{{\bf v}}
\newcommand{\xv}{{\bf x}}
\newcommand{\yv}{{\bf y}}

\newcommand{\zerov}{{\bf 0}}

\newcommand{\Am}{{\bf A}}
\newcommand{\Bm}{{\bf B}}

\newcommand{\Dm}{{\bf D}}
\newcommand{\Em}{{\bf E}}
\newcommand{\Fm}{{\bf F}}
\newcommand{\Gm}{{\bf G}}
\newcommand{\Hm}{{\bf H}}
\newcommand{\Id}{{\bf I}}

\newcommand{\Pm}{{\bf P}}
\newcommand{\Qm}{{\bf Q}}

\newcommand{\Sm}{{\bf S}}

\newcommand{\Um}{{\bf U}}

\newcommand{\Vm}{{\bf V}}

\newcommand{\Ym}{{\bf Y}}

\newcommand{\Cc}{{\cal C}}

\newcommand{\Nc}{{\cal N}}

\newcommand{\Rc}{{\cal R}}

\newcommand{\Tc}{{\cal T}}

\newcommand{\etav}{\hbox{\boldmath$\eta$}}

\newcommand{\nuv}{\hbox{\boldmath$\nu$}}

\newcommand{\Gammam}{\hbox{\boldmath$\Gamma$}}
\newcommand{\Lambdam}{\hbox{\boldmath$\Lambda$}}

\newcommand{\Psim}{\hbox{\boldmath$\Psi$}}

\newcommand{\diag}{{\hbox{diag}}}

\newcommand{\trace}{{\hbox{tr}}}

\newcommand{\herm}{^{\text{H}}}

\graphicspath{{figures/}}

\begin{document}


\title{Vandermonde-subspace Frequency Division Multiplexing for Two-Tiered Cognitive Radio Networks}
\author{Leonardo S. Cardoso, Mari Kobayashi, Francisco Rodrigo P. Cavalcanti and M\'erouane Debbah

\thanks{This work is partly funded by the CAPES-COFECUB and the Alcatel-Lucent, through the Alcatel-Lucent Chair on Flexible Radio.}
\thanks{Leonardo S. Cardoso is with the CITI lab, INSA de Lyon, France e-mail: leonardo.cardoso@insa-lyon.fr}
\thanks{Mari Kobayashi is with Dept. de T\'el\'ecommunications, Sup\'elec, Gif-sur-Yvette, France e-mail: mari.kobayashi@supelec.fr}
\thanks{F. Rodrigo. P. Cavalcanti is with GTEL - Dept. Teleinform\'atica, Universidade Federal do Cear\'a, Fortaleza, Cear\'a, Brazil e-mail: rodrigo@gtel.ufc.br}
\thanks{M\'erouane Debbah is with Alcatel-Lucent Chair on Flexible Radio, Sup\'elec, Gif-sur-Yvette, France e-mail: merouane.debbah@supelec.fr}

}

\maketitle

\begin{abstract}
Vandermonde-subspace frequency division multiplexing (VFDM) is an overlay spectrum sharing technique for cognitive radio. VFDM makes use of a precoder based on a Vandermonde structure to transmit information over a secondary system, while keeping an orthogonal frequency division multiplexing (OFDM)-based primary system interference-free. To do so, VFDM exploits frequency selectivity and the use of cyclic prefixes by the primary system. Herein, a global view of VFDM is presented, including also practical aspects such as linear receivers and the impact of channel estimation. We show that VFDM provides a spectral efficiency increase of up to 1~bps/Hz over cognitive radio systems based on unused band detection. We also present some key design parameters for its future implementation and a feasible channel estimation protocol. Finally we show that, even when some of the theoretical assumptions are relaxed, VFDM provides non-negligible rates while protecting the primary system.
\end{abstract}


\thispagestyle{empty}

\section{Introduction}

A new trend in cellular communications is currently on the rise: the deployment of smaller base stations alongside macrocell ones to aid in capacity and coverage \cite{arti:hoydis2011II,arti:chandrasekhar2008}. Differently from microcells~\cite{inpr:kishore2002}, or multi-access networks \cite{inpr:yilmaz2005}, the current trend proposes smaller base stations, called femto-~\cite{arti:chandrasekhar2008} or small-cells~\cite{arti:hoydis2011II} which are user deployed in a uncontrolled manner and need to intelligently adapt to coexist with macro base stations. Furthermore it aims to reduce the band footprint and raise the spectral efficiency by promoting a re-use of the same band and technology used by the cellular network. The problem with this paradigm is that, unmanaged, femto-cells might generate unbearable amounts of interference to macrocell communications.

The easiest way two tier networks can manage interference is by minimizing the overlap of resources between the tiers, like currently proposed in the state of the art of orthogonal frequency division multiple access (OFDMA) wireless system designs (i.e. \cite{arti:lopez-perez2009, inpr:andrews2010}). Nevertheless, separating the resources (time, frequency or power) between the two tiers limits the maximum spectral efficiency each tier can attain. Diverse means of achieving coexistence between tiers include power control \cite{arti:chandrasekhar2009}, relay spectrum sharing~\cite{arti:han2009}, beamforming~\cite{arti:islam2008} and interference alignment~\cite{inpr:shen2008}.

A whole set of  promising techniques span from the cognitive radio and dynamic spectrum access (DSA)~\cite{arti:goldsmith2009,arti:akyildiz2006} ideas. By framing the two-tiered system into the cognitive radio perspective, i.e., macro base stations and its users, acting as the primary system, are protected from interference while, femto-cells and its users, acting as the secondary system, can accept interference from the macro base stations, we can exploit the awareness of the secondary system and its cognitive capabilities to deal with the interference issue. In DSA overlay, knowledge of the primary system's characteristics are obtained in order to actively mitigate the interference from the secondary system. Dirty paper coding (DPC) \cite{arti:devroye2006}, opportunistic interference alignment \cite{arti:perlaza2009} and spectrum shaping~\cite{arti:zhang2010I} are examples of DSA overlay. 

In a previous work \cite{conf:cardoso2008}, we introduced Vandermonde frequency division multiplexing (VFDM), an overlay DSA technique for cognitive radio networks that relies on the frequency dimension. This technique is based on a linear precoder that allows a secondary transmitter to precode its signal on the null-space of the interfering channel, thereby incurring zero interference at the primary receiver. VFDM exploits the unused resources created by frequency selectivity and the use of guard symbols in block transmission systems at the primary system, such as orthogonal frequency division multiplexing (OFDM). Different from some of the underlay studied techniques, which limit the maximum power used by the secondary system~\cite{arti:kolodzy2006}, VFDM ensures interference cancelation irrespective of the primary and secondary system's transmitted data or power allocation.  

In this work, VFDM is seen from a more general perspective. Unlike our previous works, we show that there's a whole family of possible precoders and we define a way to obtain any precoder within this set. We also show that the choice of the precoder can be made in a way  to optimize for a better spectral efficiency at the secondary system. We benchmarked VFDM against one widely accepted solution for this kind of scenarios (spectrum sensing based resource partitioning and OFDM based interference alignment) showing we can indeed exploit the unused resources of the primary system to enhance the system spectral efficiency by at least 0.5 bps/Hz. Then, the spectral efficiency of VFDM is analyzed under the assumption of perfect channel state information (CSI) at the secondary transmitter and receiver. As seen in previous works, VFDM is highly dependent on perfect channel state information at the transmitter (CSIT) to be able to achieve zero interference at the primary system. In practice, perfect CSIT is not possible and an error-prone version of the CSIT is generally obtained through channel estimation, and used during the coherence time. However, to obtain a good channel estimation a large amount of training symbols need to be transmitted, cannibalizing on the actual data transmission time. In this work we also study the best tradeoff of the amount of training versus data symbols in the performance of VFDM. Furthermore, our findings reflect the behavior of any null-space precoder for this kind of problem. To better understand this training-to-data tradeoff and the impact of bad CSIT knowledge in VFDM's performance, a practical channel estimation protocol for VFDM is proposed. We show that, even under imperfect CSI at the transmitter and receiver, VFDM allows for meaningful rates to be achieved at the secondary system with no change to the primary system's technology and at the sole cost of added intelligence and awareness at the secondary system.

This work is organized as follows. In the next section, we introduce the system model and problem, assumed throughout this work. Then, a more general Vandermonde-subspace precoder design is introduced in Sec. \ref{sec:vfdm_precoder}. In Sec. \ref{sec:opt_receiver}, we analyze the achievable rates of VFDM. A channel estimation procedure is presented in Sec. \ref{sec:channel_est}. All of these findings are illustrated with some numerical results in Sec. \ref{sec:num_examples}. Finally, the concluding remarks and future research directions are given in Sec. \ref{sec:conclusions}.

In this work we have adopted the mathematical notation as described in the following. A lower case italic symbol (ex. $b$) denotes a scalar value, a lower case bold symbol (ex. $\bv$) denotes a vector, an upper case bold symbol (ex. $\Bm$) denotes a matrix and an upper case bold symbol within square brackets (ex. $[\Bm]_{mn}$) denotes a matrix element at the $m^{\text{th}}$ row and the $n^{\text{th}}$ column. An $\Id_{N}$ denotes the identity matrix of size $N$. The transpose conjugate operator on a matrix is denoted by the $^\text{H}$ superscript (ex. $\Fm^{\text{H}}$). All vectors are columns, unless otherwise stated.

\section{System Model}\label{sec:system_model}

\begin{figure}[htbp]
   \centering
   \def\svgwidth{0.8\columnwidth}
   \input{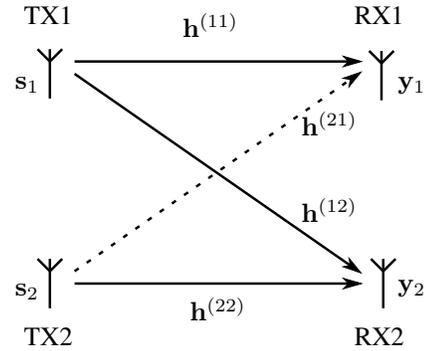}
   \caption{Cognitive interference channel model}
   \label{fig:scenario}
\end{figure}

To understand the overlay DSA problem, consider the cognitive interference channel scenario depicted in Fig.~\ref{fig:scenario}, where all transmitters and receivers have a single antenna. The cognitive interference channel is characterized by a primary system (TX1~|~RX1) that communicates a message $\sv_{1}$ over a licensed band and a secondary system (TX2 | RX2) that exploits the band opportunistically to communicate its own message $\sv_{2}$, while avoiding harmful interference to the primary receiver. The primary system, being the legal licensee of the band, does not need to avoid interference to the secondary system and is completely unaware of the latter.

In this work, we consider no cooperation between the primary and secondary systems. If both primary and secondary systems can fully cooperate by sharing information through an unlimited backhaul, then primary and secondary can be considered as part of a network multiple input multiple output (MIMO) system. In this case, if all the messages ($\sv_{1}$ and $\sv_{2}$) are known prior to transmission at all the transmitters (TX1 and TX2), then the cognitive interference channel can be generalized to a 2~$\times$~2 MIMO broadcast channel for which DPC has been shown to be capacity achieving~\cite{arti:weingarten2006}. In this case, interference is suppressed on both cross links ($\hv^{(12)}$ and $\hv^{(21)}$). The asymmetric case, particular for the cognitive interference channel of Fig.~\ref{fig:scenario}, has also been shown to capacity achieving by DPC \cite{arti:devroye2006}. Non-optimally, a zero forcing (ZF) scheme~\cite{arti:caire2003} can be used in the fully cooperative network MIMO case to null the interference towards the unintended receivers. Finally, if all transmitters know the channels to all the receivers and multiple antennas are adopted, than interference alignment (IA)~\cite{arti:cadambe2008} becomes an interesting candidate. IA starts providing interesting gains, in terms of degrees of freedom, when the system size is larger than 3x3~\cite{arti:cadambe2008}. A proposal for an OFDM based IA has been made earlier~\cite{inpr:shen2008}, resembling more the cognitive interference channel scenario, target of this work. However, it requires cooperation between radios, since knowledge of all the interference subspaces at the receivers is needed. Furthermore, the addition of a zero forcing interference nulling matrix imposes performance penalties to the received signal. 

For the specific purposes of this work, we consider the frequency selective version of the cognitive interference channel of Fig.~\ref{fig:scenario}, comprised of $L+1$ tap frequency selective channels between transmitter $i$ and receiver $j$, $\hv^{(ij)}$. The channels entries are unit-norm, independent and identically distributed (i.i.d.), complex circularly symmetric and Gaussian ${\Cc}\Nc(0,\Id_{L+1}/(L+1))$. Furthermore, the channels are i.i.d. over any pair $i,j$. 

We consider that both the primary and secondary systems employ an $N+L$ size block transmission in order to cope with block-interference. At the primary system, classical OFDM with $N$ subcarriers and a cyclic prefix of size $L$ is used. The choice of OFDM for the primary system is merely to provide a practical setting, but the results presented in this work can be extended to any block transmission system that employs guard symbols (discarded at the reception). For the secondary system, an OFDM-like block transmission scheme is adopted, where the leading $L$ symbols are also discarded. We assume time division duplex (TDD) transmissions (where channel reciprocity can be exploited), and that proper RF calibration~\cite{inpr:guillaud2005} or transceiver construction ensures reciprocity at base-band level. We also assume that the signals are synchronized at the reception (both at the primary and secondary systems) at symbol-level. At this point of the work, let us also assume perfect knowledge of the CSI (further in this work, this assumption will be relaxed). Then, the received signals at both the primary and secondary receivers are given by
\begin{eqnarray} 
    \yv_1 &=& \Fm \left(\Tc(\hv^{(11)})\xv_1 + \Tc(\hv^{(21)})\xv_2+ \nv_1 \right)\label{eq:y1}\\
    \yv_2 &=& \Fm \left(\Tc(\hv^{(22)}) \xv_2 + \Tc(\hv^{(12)}) \xv_1+ \nv_2 \right),\label{eq:y2}
\end{eqnarray}
where $\Tc(\hv^{(ij)}) \in {\Cc}^{N\times(N+L)}$ is matrix with a Toeplitz structure constructed from the channel's coefficients given by
\begin{eqnarray*}
\Tc(\hv^{(ij)}) =
\left[ \begin{array}{cccccc}
h_{L}^{(ij)} & \cdots & h_0^{(ij)} & 0 & \cdots & 0 \\
0 & \ddots &  & \ddots & \ddots & \vdots \\
\vdots & \ddots & \ddots &  & \ddots & 0 \\
0 & \cdots & 0 & h_{L}^{(ij)} & \cdots & h_0^{(ij)} \\
\end{array} \right],
\end{eqnarray*}
$\Fm \in \Cc^{N\times N}$ is a unitary discrete Fourier transform (DFT) matrix with $[\Fm]_{k+1, l+1}=\frac{1}{\sqrt{N}}e^{-i2\pi \frac{kl}{N}}$ for $k,l = 0,\dots,N-1$, and $\xv_i$ denotes the transmit vector of user $i$ of size $N+L$ subject to the individual power constraint given by
\begin{equation}\label{eq:pconst}
 \trace(\EE[\xv_i\xv_i^{\text{H}}]) \leq (N+L) P_i
\end{equation}
and $\nv_i\sim\Cc\Nc(0, \sigma_n^{2}\Id_N)$ is an $N$-sized additive white gaussian noise (AWGN) noise vector. The transmit power per symbol is $P_{i}$. For the primary system, we consider OFDM-modulated symbols
\begin{equation}\label{eq:x1}
  \xv_1 = \Am \Fm^{-1} \sv_1
\end{equation}
where $\Am$ is a $(N+L)\times N$ a cyclic prefix precoding matrix that appends the last $L$ entries of $\Fm^{-1} \sv_1$ and $\sv_1$ is a symbol vector of size $N$ and unitary norm. Regarding the secondary user, the transmit vector is given by 
\begin{equation}\label{eq:x2}
  \xv_2 =\Em\sv_2,
\end{equation}
where $\Em \in {\Cc}^{(N+L)\times L}$ is a linear precoder and $\sv_2$ is a unitary norm symbol vector. $\Em$ will be presented in the next section.

As previoulsy stated, the secondary system tries to cancel its interference to the primary one, while the primary system remains oblivious to the presence of the secondary one. This is effectively achieved when
\begin{equation}\label{eq:zero_interf}
\Fm\Tc(\hv^{(21)}) \Em \sv_2 = \zerov ~ \forall ~ \sv_2,
\end{equation}
meaning that $\Hm_{21} = \Fm\Tc(\hv^{(21)}) \Em = \zerov$, obtained by substituting (\ref{eq:x1}) and (\ref{eq:x2}) into (\ref{eq:y1}) and making the interference part equal to zero. The signal received at the primary system becomes
\begin{equation} \label{eq:Y1}
    \yv_1 = \Hm_{\text{11}} \sv_1 + \nuv_1,
\end{equation}
where $\Hm_{\text{11}} = \Fm\Tc(\hv^{(11)})\Am\Fm^{-1}$ is an $N \times N$ diagonal overall channel matrix for the primary system and $\nuv_1$ the Fourier transform of the noise $\nv_1$, has the same statistics as $\nv_1$.

The primary system does not cooperate with the secondary one, which performs single user decoding at the secondary receiver. Therefore, we take
\begin{equation}\label{eq:eta}
\etav = \Hm_{\text{12}} \sv_1 + \nuv_2,
\end{equation}
as the interference plus noise component, obtained when substituting (\ref{eq:x1}) and (\ref{eq:x2}) into (\ref{eq:y2}), where $\Hm_{\text{12}}=\Fm \Tc(\hv^{(12)}) \Am \Fm^{-1}$ is an $N \times N$ diagonal overall channel matrix for the primary system and $\nuv_2$ the Fourier transform of the noise $\nv_2$, has the same statistics as $\nv_2$. The use of a DFT and the removal of the leading $L$ symbols at the secondary receiver makes it possible to consider a diagonal $\Hm_{\text{12}}$, which in turn, allows a simplification of the subsequent analysis w.r.t. $\etav$. The signal received at the secondary system becomes
\begin{equation}\label{eq:Y2}
 \yv_2 = \Hm_{\text{22}} \sv_2 +  \etav,
\end{equation}
where $\Hm_{\text{22}}=\Fm \Tc(\hv^{(22)})\Em$ denotes the overall $N\times L$ secondary channel. Finally, from (\ref{eq:Y1}) and (\ref{eq:Y2}), we remark that VFDM successfully converts the frequency-selective interference channel (\ref{eq:y1}) and (\ref{eq:y2}) (or X interference channel) into an one-side vector interference channel (or Z interference channel) where the primary receiver sees interference-free $N$ parallel channels. Even so, the secondary receiver sees interference from the primary transmitter. 

\section{Precoder Design}\label{sec:vfdm_precoder}

To achieve its best performance, the secondary system must be designed with two goals in mind: 1) maximize the achievable rate at the secondary system; 2) enforce the interference protection at the primary system. In mathematical terms, this is equivalent to finding a $\Sm_{2}$, $\Em$ pair that simultaneously solves the optimization problem
\begin{scriptsize}
\begin{eqnarray}\label{eq:general_opt}
    \max_{\Sm_{\text{2}},\Em} && \left( \frac{1}{N+L} \log_2\left|\Id_N + \Sm_{\eta}^{-1/2}\Fm \Tc(\hv^{(22)})\Em \Sm_2 \Em\herm \Tc(\hv^{(22)})\herm \Fm\herm  \Sm_{\eta}^{-\text{H}/2}\right|\right)  \nonumber \\ 	
\text{s.t.} && \left\{ \begin{array}{ccl}
 \trace(\Em\herm \Em\Sm_{2}) & \leq & (N+L)P_{2}\\
 \Tc(\hv^{(21)}) \Em & = & \zerov.
 \end{array} \right.,
\end{eqnarray}
\end{scriptsize}
where $\Sm_{2}$ is the covariance matrix of $\sv_2$, $\Sm_{\eta} =  \Hm_{\text{12}} \Sm_1 \Hm_{\text{12}}\herm + \sigma_{\text{n}}^{2}\Id_N$ is the covariance matrix of $\etav$, the first restriction comes from (\ref{eq:pconst}) and the second restriction comes from (\ref{eq:zero_interf}). We approximate $\etav$ to a zero-mean Gaussian random vector. Note that, the objective function in (\ref{eq:general_opt}) does not take into consideration the rate at the primary system. This is the case since, by guaranteeing zero interference from the secondary system, the primary system can achieve maximum capacity through the optimization of its own input power allocation though classical water filling~\cite{inpr:rhee2000}.

A closed form solution to (\ref{eq:general_opt}) is not known. A similar problem has been addressed in \cite{inpr:huh2009} (based on \cite{arti:wiesel2008}) where a numerical solution, using convex relaxation and generalized inverses, is used to find the optimal steering vectors and transmit powers. In this work, by exploiting the fact that the second restriction gives a clue on the design constraints of $\Em$, a solution is proposed. Due to the particular structure of $\Tc(\hv^{(21)})$, it is not difficult to show that a matrix $\Em$, capable of yielding $\Tc(\hv^{(21)}) \Em = \zerov$, has to evaluate the polynomial
\begin{displaymath}
S(z) = \sum_{i=0}^L h_i^{(21)} z^{L-i},
\end{displaymath}
at its roots $\{a_l,\dots,a_L\}$. Interestingly, the Vandermonde matrix is known for its property to evaluate a polynomial at certain values \cite{book:golub1996}. In fact, it is straightforward to see that
\begin{eqnarray}
	{\Vm} = \left[ \begin{array}{cccccc}
		1 & \cdots & 1 \\ 
		a_1 & \cdots & a_{L} \\
		a^2_1 & \cdots & a^2_{L} \\
		\vdots & & \vdots \\
		a^{N+L-1}_1 & \cdots & a^{N+L-1}_{L} \\
	\end{array} \right], 
\end{eqnarray}
defines the null-space of $\Tc(\hv^{(21)})$ and, without loss of generality, we can further choose the columns of $\Em$ as any linear combination based on the $L$ columns of $\Vm$, such that
\begin{equation}\label{eq:columns}
[\Em]_{k} = \sum_{l=1}^{L} [\Gammam]_{k,l}\vv_{l}
\end{equation}
where $[\Gammam]_{k,l}$ is the $(k^{\text{th}},l^{\text{th}})$ element of $\Gammam \in \Rc^{L \times L}$, a coefficient matrix. Translating \eqref{eq:columns} into a matricial form we finally obtain $\Em$ as 
\begin{equation}\label{eq:v}
\Em \triangleq \Vm \Gammam.
\end{equation}
Equation~\eqref{eq:v} defines a set of suitable precoders that belong to the null-space of $\Tc(\hv^{(21)})$. Note that any a precoder is obtained by tuning the coefficients of $\Gammam$ accordingly. Since the precoder is based on a Vandermonde generated subspace and its orthogonality w.r.t. $\Tc(\hv^{(21)})$ enables the users to transmit simultaneously over the same frequency band, we have decided to name this scheme {\it Vandermonde-subspace Frequency Division Multiplexing} (VFDM).

In practice, $\Em$ is constructed by selecting $\Gammam$ such that a set of $L$ orthonormal columns, that lie inside of the Vandermonde-subspace of $\Tc(\hv^{(21)})$, is found. This can be accomplished by finding the QR decomposition \cite{book:meyer2000} of $\Vm = \Em\Gammam^{-1}$, where $\Gammam^{-1}$ is an upper triangular matrix and $\Em$ is orthonormal. Since $\Vm$ will be nonsingular with a high probability, $\Em$ is unique and can be numerically obtained by performing a Gram-Schmidt process \cite{book:meyer2000} on the columns of $\Vm$. Another way to construct $\Em$ is by singular value decomposition (SVD) of $\Tc(\hv^{(21)})$ \cite{arti:kobayashi2009}, where if $\Tc(\hv^{(21)}) = \Um_{\Tc} \Lambdam_{\Tc} \Vm_{\Tc}^{\text{H}}$, then 
\begin{displaymath}
\Em = \left[~ \vv_{\Tc N} ~|~ \cdots ~|~ \vv_{\Tc(N+L)-1} ~|~ \vv_{\Tc N+L} ~\right], 
\end{displaymath}
and $\Vm_{\Tc}$ has the form $\left[ ~ \vv_{\Tc1} ~|~\vv_{\Tc2} ~|~ \cdots  ~|~ \vv_{\Tc N+L} ~ \right]$.

As hinted in the above development, the structure of the $\Em$ precoder is dependent on the number of taps $L$. Indeed, if a number of taps $l < L$ is used, a reduced number of degrees of freedom can be used to transmit useful symbols from the primary transmitter, limiting the effectiveness of the technique.

\section{Achievable Rates}\label{sec:opt_receiver}

In order to determine the achievable rates at the secondary system, we introduce a simplification: instead of trying to solve   (\ref{eq:general_opt}) over both $\Sm_2$ and $\Em$, we choose a $\Gammam$ matrix that provides a well-behaved $\Em$ with orthonormal columns, and optimize only on $\Sm_2$. The optimization problem now becomes
%
%
\begin{eqnarray}\label{eq:new_opt}
    \hspace{-2.5cm}\max_{\Sm_{\text{2}}} & & \left( \frac{1}{N+L} \log_2\left|\Id_N + \Sm_{\eta}^{-1/2}\Hm_{\text{22}}\Sm_2 \Hm_{\text{22}}\herm \Sm_{\eta}^{-\text{H}/2}\right|\right) \\ 	
\text{s.t.} & & \trace(\Em\herm \Em\Sm_{2}) \leq (N+L)P_{2}, \label{eq:linear_const}
\end{eqnarray}
%
%
where we have dropped the second restriction since it becomes implicit from the precoder design.

At this point some remarks are of order. Since the channels $\hv^{(21)}$ and $\hv^{(22)}$ are statistically independent, the probability that $\Tc(\hv^{(21)})$ and $\Tc(\hv^{(22)})$ have the same null-space is zero. Hence, we can expect that the secondary user's symbols $\sv_2$ will be transmitted reliably. Furthermore, since the precoder does not depend on the transmitted symbols and due to the orthogonality between the channel and the precoder, the orthogonality condition (\ref{eq:zero_interf}) always holds irrespectively of the secondary system's input power $P_2$ and its link. Clearly, perfect knowledge of the $\hv^{(21)}$ CSI is required at the secondary transmitter in order to adapt the precoder to the channel fluctuations. In addition to that, perfect knowledge of the interference plus noise covariance $\Sm_{\eta}$ of the secondary receiver is also required at the secondary transmitter. Practical aspects on how to perform channel estimation is given in detail in Sec. \ref{sec:channel_est}.

The new optimization problem in (\ref{eq:new_opt}) is also convex, but the presence of the term ${\Em}\herm \Em$ in the constraint (\ref{eq:linear_const}) requires a prior manipulation step in order to obtain a water-filling solution. Let us initially define
\begin{displaymath}
\Gm = \Sm_{\eta}^{-1/2}\Hm_{\text{22}},
\end{displaymath}
with $\Gm \in \Cc^{N \times L}$ to be an equivalent channel. We then take the SVD of the equivalent channel $\Gm = \Um_{\text{G}} {\Lambdam_{\text{G}}}^{1/2} {\Vm_{\text{G}}}\herm$, where $\Um_{\text{G}} \in \Cc^{N \times N}$ and $\Vm_{\text{G}} \in \Cc^{L \times L}$ are unitary matrices and $\Lambdam_{\text{G}} \in \Rc^{N \times L}$ contains a top diagonal with the $r \leq L$  eigenvalues of $\Gm\herm\Gm$, with $[ \Lambdam_{\text{G}} ]_{i,i} \geq 0$. Finally, we let $\Sm_2 = \Vm_{\text{G}} \Dm_2 {\Vm_{\text{G}}}\herm$ with $\Dm_2 \in \Rc^{L \times L}$ being $\diag[d_1, \cdots, d_L ]$, and $\Qm = {\Vm_{\text{G}}}\herm{\Em}\herm \Em{\Vm_{\text{G}}}$. Using these new definitions (\ref{eq:new_opt}) and (\ref{eq:linear_const}) become
\begin{eqnarray}
	\max_{\Pm_{2}} & & \left( \frac{1}{N+L} \log\left|\Id_N + \Um_{\text{G}} \Lambdam_{\text{G}}^{1/2}\Dm_2 \Lambdam_{\text{G}}^{\text{H}/2} \Um_{\text{G}}\herm \right|\right) \nonumber \\ 	
	\text{s.t.} & & \trace(\Qm\Dm_{2}) \leq (N+L)P_{2}, \nonumber
\end{eqnarray}
which can be further rewritten in scalar form as
\begin{eqnarray}\label{eq:final_opt}
    \max_{d_i} & & \sum_{i=1}^{L}\log_{2}(1 + d_{i}[\Lambdam_{\text{G}}]_{i,i})  \\
    \text{s.t.} & & \sum_{i=1}^{L} d_{i} [\Qm]_{i,i} \leq (N+L)P_{2}. \nonumber
\end{eqnarray}
The optimization problem in its new form (\ref{eq:final_opt}) can be easily solved using the Karush-Kuhn-Tucker (KKT) conditions which lead to the classical water-filling solution \cite{book:tse2005}. The solution to (\ref{eq:final_opt}) is given by $\Sm_{\text{2}} = \Vm_{\text{G}} \Dm_2 \Vm_{\text{G}}^{\text{H}}$, where the $i^{\text{th}}$ component of the matrix $\Dm$ is the weighted water-filling solution given by
\begin{equation}\label{eq:wf}
d_i = \left[ \frac{\mu}{[\Qm]_{i,i}} -\frac{1}{[\Lambdam_{\text{g}}]_{i,i}}\right]_+,
\end{equation}
where $\mu$, known as the ``water level'', is determined to fulfill the total power constraint $(N+L)P_{\text{2}}$. Since we have chosen $\Gammam$ such that $\Em$ is orthonormal, it follows that $\forall i, [\Qm]_{i, i} = 1$, and therefore, the maximum achievable spectrum efficiency for the secondary system is finally given by
\begin{equation}\label{eq:rate}
    R_{\text{opt}} = \frac{1}{N+L}\sum_{i=1}^{L} \log_{2}(1 + d_{i}[\Lambdam_{\text{G}}]_{i,i}).
\end{equation}

\section{Channel Estimation Protocol}\label{sec:channel_est}

As seen previously, VFDM requires the CSIT in order to create the $\Em$ precoder. To devise a practical secondary VFDM system, a channel estimation protocol  needs to be designed, taking into consideration prior knowledge of the primary system and its own channel estimation procedure. 

One stringent constraint for wireless systems is the channel coherence time $T$. All channel estimations and the transmission itself need to take place before the channel state changes. Using training sequences to perform channel estimation consume valuable resources that sacrifice spectral efficiency. Nevertheless, the larger the amount of symbols used for channel estimation, the better the estimate, yielding higher rates. Therefore, the design of an efficient channel estimation protocol is of utmost importance. 

The objectives of this section are twofold: devise a channel estimation procedure and evaluate the impact of training versus transmit symbols on the performance of VFDM. 

\subsection{Primary System}

We start off by assuming that the primary receiver transmits training symbols on the uplink back to the primary transmitter. This prior step is justified by the necessity to adapt the primary transmitter's own downlink power allocation with respect to the CSI of $\hv^{(11)}$. Throughout this section, a least squares (LS) channel estimation procedure is adopted~\cite{arti:biguesh2006},~\cite{inpr:li2000}. In the uplink, $\Psim_{\text{1}} \in \Cc^{N \times \tau_{\text{u}}}$ are orthogonal Fourier-based training symbols sent during a time $\tau_{\text{u}} \geq N+L$, that depends on the amount of symbols required for the channel estimation. After that, the primary transmitter sends back the same training symbols $\Psim_{\text{1}}$ during a time $\tau_{\text{d}} \geq N+L$, used by the primary receiver to estimate $\hv^{(11)}$, this time for equalization purposes. The total channel estimation time for the primary system $\tau = \tau_{\text{u}}+\tau_{\text{d}}$.

\subsection{Secondary System}

During $\tau_{\text{u}}$, the secondary transmitter taps into the primary system's uplink training transmission, receiving
\begin{displaymath}
\Ym_{2\text{u}} = \Hm_{21}' \Psim_{\text{1}} + \mathbf{\Upsilon}_{2\text{u}},
\end{displaymath}
where $\Ym_{2\text{u}} \in \Cc^{N \times \tau_{\text{u}}}$ is a matrix of received OFDM symbols during a time $\tau_{\text{u}}$, $\Hm_{21}'=\Fm\Tc(\hv^{(21)})\Am\Fm^{-1}$ is a diagonal overall channel matrix from the primary receiver to the secondary transmitter and $\mathbf{\Upsilon}_{2\text{u}} \in \Cc^{N \times \tau_{\text{u}}}$ is the overall received noise matrix over time $\tau_{\text{u}}$. We remark that $\Hm_{21}'$ is essentially different from $\Hm_{21}$, since the latter is the (not diagonal) overall channel including the precoder $\Em$. The estimate $\hat{\Hm}_{21}'$ is then obtained by
\begin{eqnarray}
\hat{\Hm}_{21}' &=& \frac{\sqrt{N}}{\tau_{\text{u}}}\Ym_{2\text{u}}\Psim_{\text{1}}\herm \nonumber \\
&=& \Hm_{21}' + \frac{\sqrt{N}}{\tau_{\text{u}}}\mathbf{\Upsilon}_{2\text{u}}\Psim_{\text{1}}\herm.\label{eq:chanest_prim}
\end{eqnarray}
The uplink channel estimation in (\ref{eq:chanest_prim}) is made possible since the secondary system possesses prior knowledge of the channel estimation procedure and training symbol structure of the primary system.

To construct $\Em$, the secondary transmitter must first convert the overall channel estimation $\hat{\Hm}_{21}'$ to the time domain $\hat{\hv}^{(21)}  =  \left[\hat{h}^{(21)}_{0},\hdots,\hat{h}^{(21)}_{L}\right]^{\text{T}}$ where
\begin{equation}\label{eq:chan_est}
\hat{h}^{(21)}_{k-1}  =  \left[\Fm^{-1} \hat{\Hm}_{\text{21}}' \right]_{k,k}~~k \in \{1,\hdots,L+1\}.
\end{equation}
$\Em$ can be finally constructed as described in Sec. \ref{sec:vfdm_precoder}, finishing the uplink channel estimation cycle.

A new channel estimation procedure is necessary on the downlink, to enable the secondary receiver to equalize the subsequent transmitted symbols. At this stage, the secondary transmitter sends precoded Fourier-based pilot symbols $\Psim_{\text{2}}$ (another set, orthogonal to  $\Psim_{\text{1}}$, such that $\Psim_{\text{1}}\Psim_{\text{2}}\herm$ = \zerov) to the secondary receiver, so it can estimate the channel $\hat{\hv}^{(22)}$. Due to the orthogonality between $\Psim_{\text{1}}$ and $\Psim_{\text{2}}$, the secondary transmission does not interfere on the primary's channel estimation and vice versa. The received training signal for the secondary user becomes
\begin{equation}
\Ym_{2\text{d}} = \Hm_{22} \Psim_{\text{2}} + \mathbf{\Upsilon}_{2\text{d}},\label{eq:y2imperf}
\end{equation}
which allows a similar channel estimation procedure as for the primary case, given by
\begin{eqnarray}
\hat{\Hm}_{22} &=& \frac{\sqrt{N}}{\tau_{1\text{d}}}\Ym_{2\text{d}}\Psim_{\text{2}}\herm \nonumber \\
&=& \Hm_{22} + \frac{\sqrt{N}}{\tau_{1\text{d}}}\mathbf{\Upsilon}_{2\text{d}}\Psim_{\text{2}}\herm.\label{eq:chanest}
\end{eqnarray}

Finally, both systems engage in the transmission phase during $T-\tau$. For the secondary system, transmission and reception is carried out as described in the previous sections. Unlike  the perfect CSI case, the received signal for the primary user in (\ref{eq:Y1}) now becomes
\begin{equation}
\yv_1 = \Hm_{\text{11}}\sv_{1} + \Hm_{\text{21}}\sv_{2} + \nuv_1,\label{eq:y1imperf}\\
\end{equation}
where $\Hm_{\text{21}}\sv_{2} \neq \zerov$. This is due to the channel estimation error in $\hat{\hv}^{(21)}$ that breaks the orthogonality between $\Tc(\hv^{(21)})$ and $\Em$.

We note that channel estimation in TDD systems are challenging, especially the ones in which an uplink estimation is used for a downlink transmission (or vice versa). The issue relies on the fact that the analog circuitry in the receiver and transmitter sections of the related radios are inherently different due to miss-calibration of simply physical deviations over time. While calibration techniques exist to account for these differences, they require coordination among radios, something which is not possible under the cognitive radio paradigm. Recently, a technique to deal with the cognitive radio calibration without coordination was proposed~\cite{inpr:kouassi2011,inpr:zayen2012}. With a some changes these techniques could be adapted for VFDM. 

\section{Numerical Examples}\label{sec:num_examples}

To illustrate the performance of VFDM, numerical results are produced through Monte Carlo based simulations. The parameters used in this section are inspired by IEEE 802.11a ~\cite{tech:ieee1999}. All channels and noise are generated according to the definitions made in Sec. \ref{sec:system_model}. Transmit powers are considered to be unitary for both primary and secondary system, and the signal to noise raio (SNR) is controlled by varying the noise variance $\sigma_{n}^{2}$. $\Em$ is generated by a Gram-Schmidt orthonormalization procedure on the columns of $\Vm$, as described in Sec. \ref{sec:vfdm_precoder}. For some of the presented results, in order to control the secondary system's performance with respect to the interference coming from the primary system,  an interference weighting factor $\alpha \in \left[0,1\right]$ has been added to (\ref{eq:eta}) such that
\begin{equation}\label{eq:weighted_eta}
\etav = \alpha\Hm_{\text{12}} \sv_1 + \nuv_2.
\end{equation}

\subsection{Achievable rates}

\begin{figure}[!h]
	\centering
	\includegraphics[width=\columnwidth]{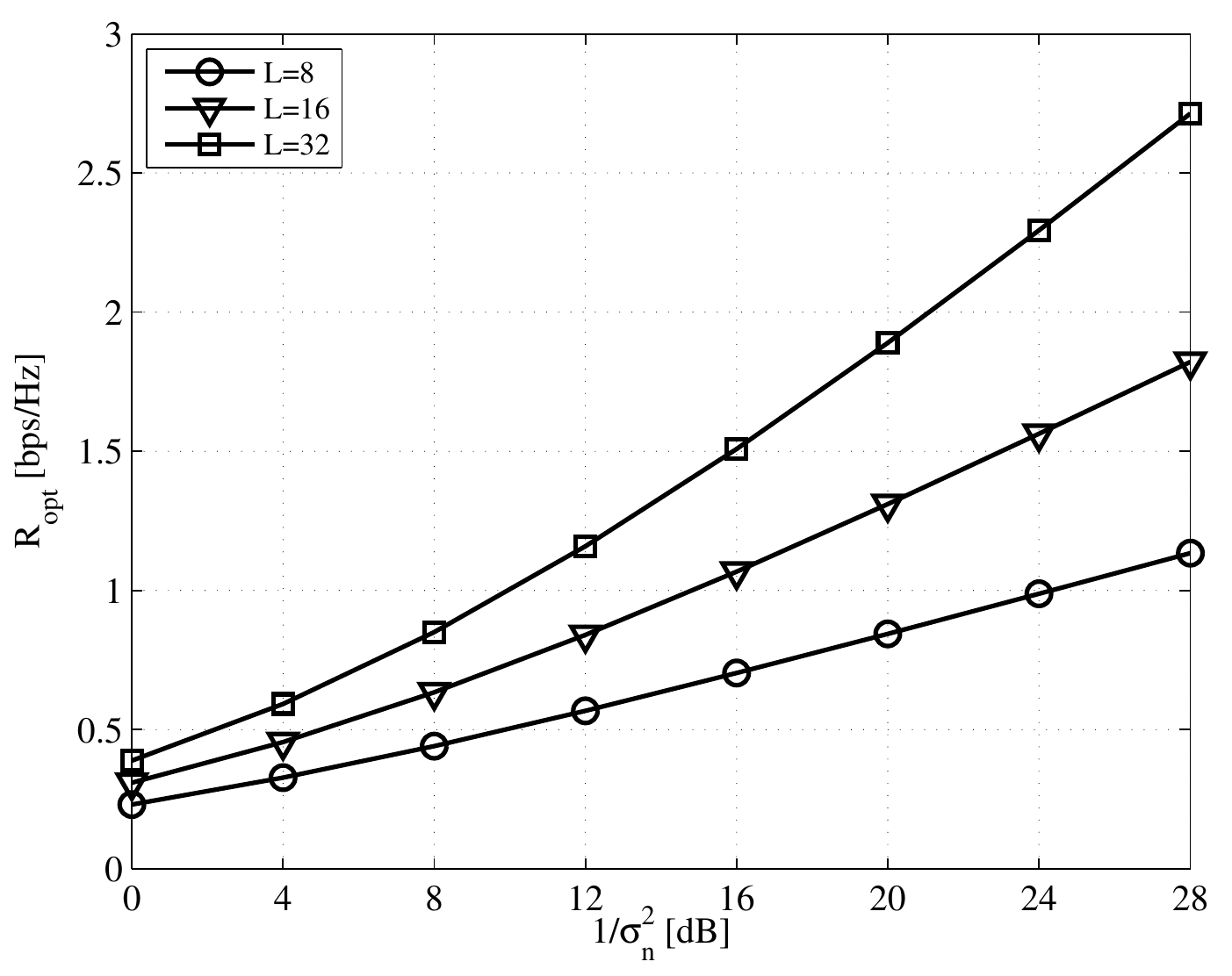} 
	\caption{VFDM's spectral efficiency for $N=64$, $L \in \{8,16,32\}$ and $\alpha$=0.}
	\label{fig:spectreff_L}
\end{figure}

In Fig.~\ref{fig:spectreff_L}, VFDM's average achievable spectrum efficiency using an optimal receiver is given for $N=64$ and three sizes of channel $L \in \{8,16,32\}$ taps. In order to isolate the performance of the secondary system, $\alpha$ is taken to be zero. The spectrum efficiency is seen to suffer a higher penalty for smaller values of $L$, since this directly translates into a smaller number of available precoding dimensions. We remark that these precoding dimensions depend only on the number of frequency selective channel taps offered by environment, which are freely exploitable. 

\begin{figure}[!h]
	\centering
	\includegraphics[width=\columnwidth]{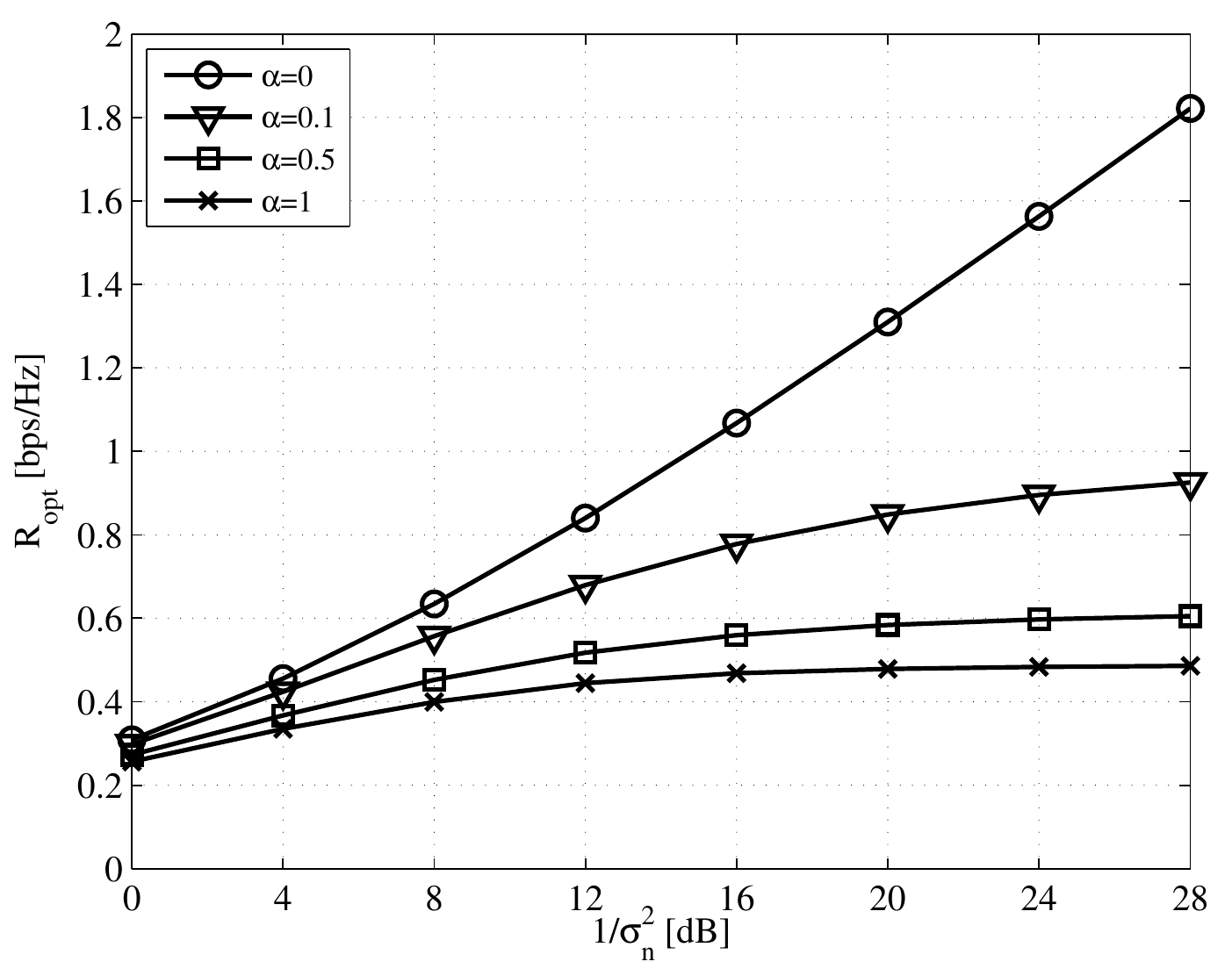} 
	\caption{VFDM's spectral efficiency for $N=64$, $L=16$ and $\alpha \in \{0,0.1,0.5,1\}$.}
	\label{fig:spectreff_alpha}
\end{figure}

In Fig.~\ref{fig:spectreff_alpha}, the effect of interference coming from the primary system on the spectral efficiency performance of VFDM is shown. As expected, the secondary system quickly becomes interference limited the higher the $\alpha$. Nevertheless, it is interesting to see that even in the worse case scenario, VFDM is still able to offer non-negligible rates.

\subsection{Comparison to Other Techniques}

\begin{figure}[!h]
	\centering
	\includegraphics[width=\columnwidth]{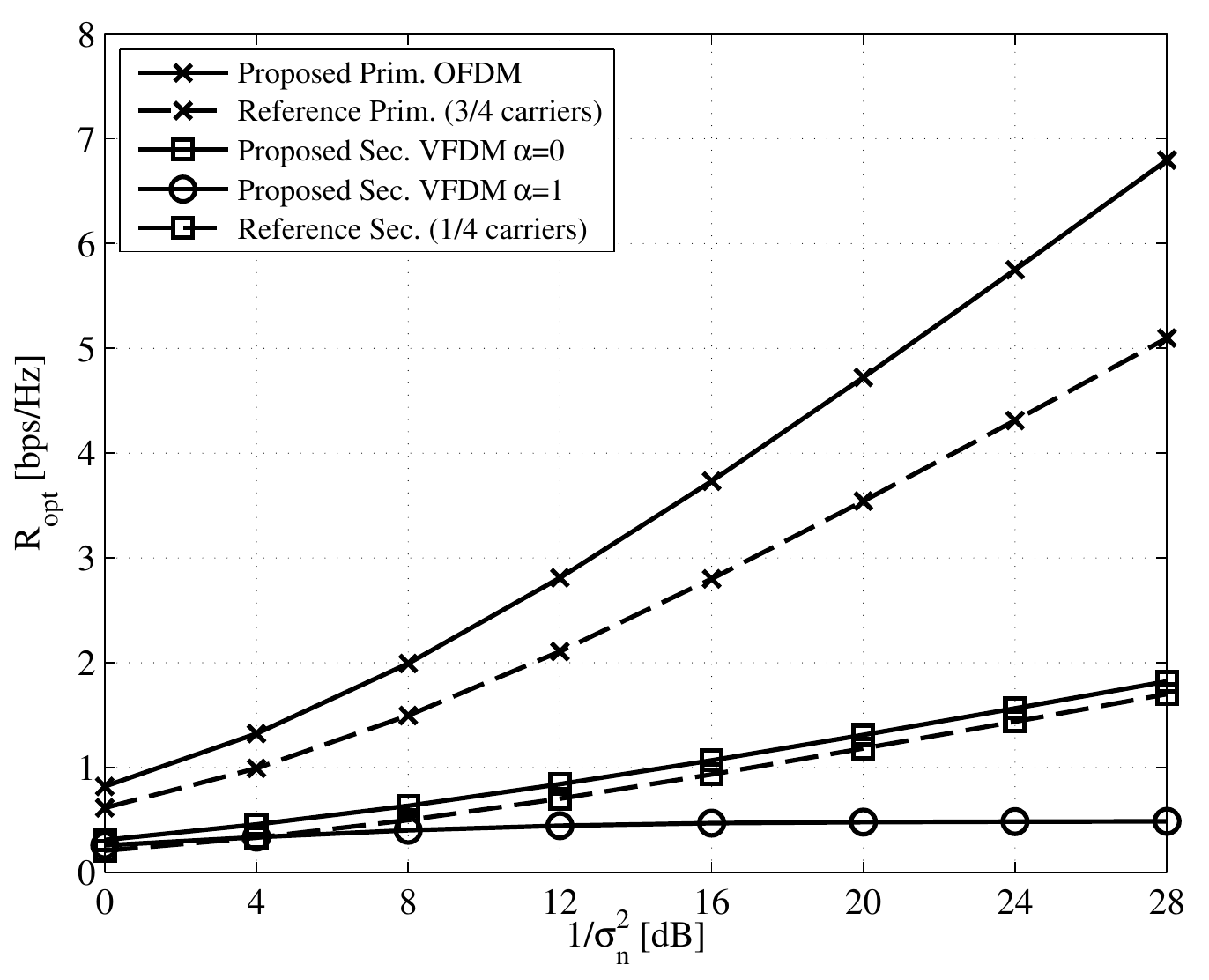} 
	\caption{VFDM's spectral efficiency for $N=64$, $L=16$ and $\alpha \in \{0,1\}$, compared to the reference primary-secondary system with divided resources.}
	\label{fig:ofdm_vs_vfdm}
\end{figure}

\begin{figure}[!h]
	\centering
	\includegraphics[width=\columnwidth]{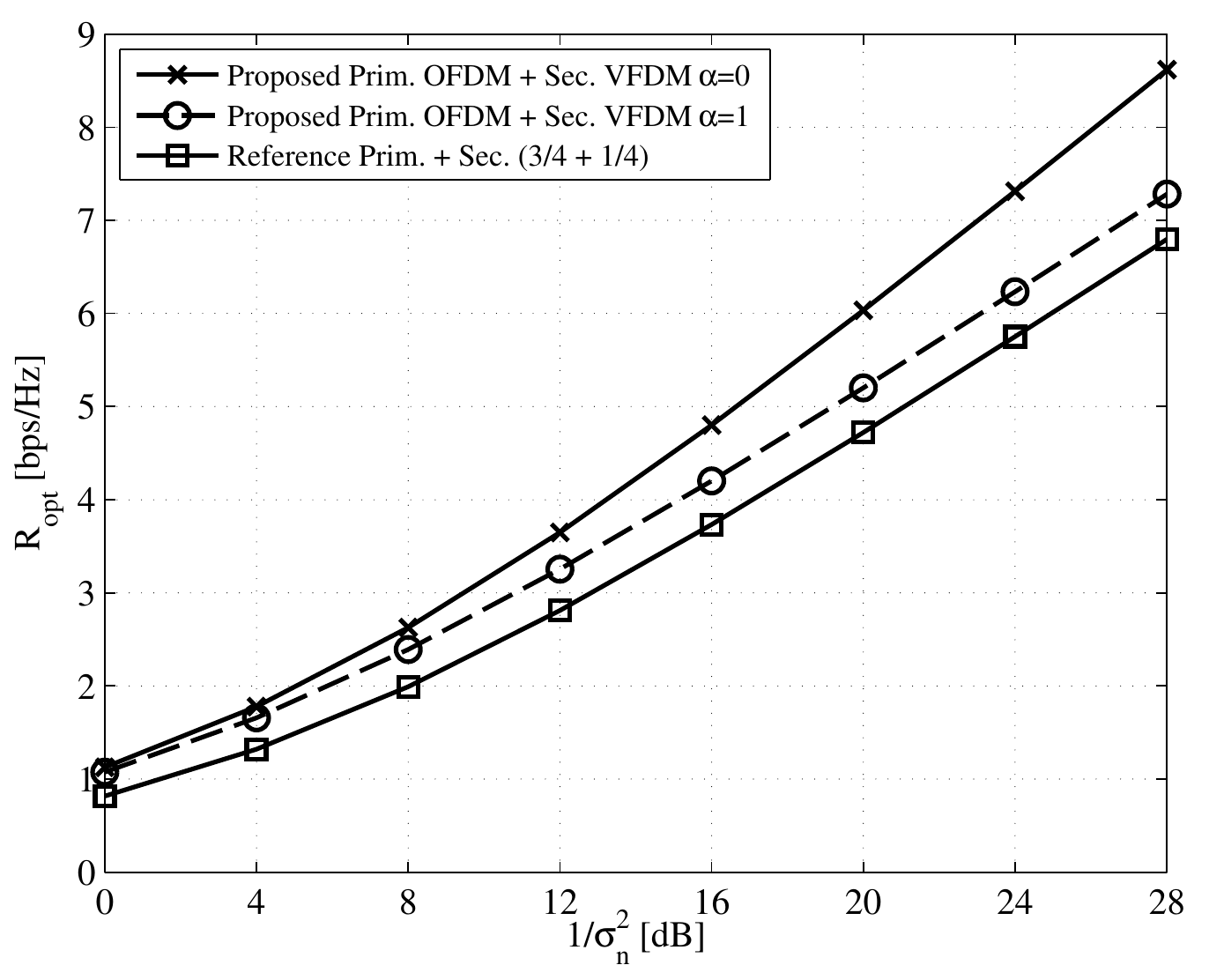} 
	\caption{VFDM's sum (primary + secondary) spectral efficiency for $N=64$, $L=16$ and $\alpha \in \{0,1\}$, compared to the reference primary-secondary system with divided resources.}
	\label{fig:ofdm_vs_vfdm_sum}
\end{figure}

As previsouly stated, one practical way of guaranteeing the coexistence between two tiers is by dividing the resources between them (e.g., \cite{arti:lopez-perez2009, inpr:andrews2010}). Multiple carriers primary systems employing this kind of coexistence technique select only a subset of channels according to specific rule, for example, channel quality indicators (CQIs). Then, by means of spectrum sensing, a secondary system can detect the free carriers and transmit at the same time on those unused frequencies. Herein, the performance of VFDM is compared with that of an OFDM-based carrier division system. In the comparison, we assume that the primary system reserves 1/4 of the carriers to the secondary system, keeping 3/4 of the carriers to itself. The remaining 1/4 of the carriers are opportunistically exploited by the reference secondary system. The 3/4|1/4 choice is made for the sake of a fair comparison, due to the fact that the simulated VFDM secondary system transmits $L=16$ symbols while the primary OFDM transmits $N=64$, hence the 1/4 proportion. Nevertheless, the results and conclusions that follow, hold for any carrier proportion. In the figures that follow, to facilitate the understanding, all labels that contain "reference" refer to the primary-secondary system pair that divide the resources, while all labels that contain "proposed" refer to the primary-secondary pair based on ODFM-VFDM. In Fig.~\ref{fig:ofdm_vs_vfdm}, we see that, the VFDM secondary system has a comparable performance to the reference secondary system. The VFDM secondary system is subject to interference from the proposed primary system and can experience a severely degraded performance for $\alpha=1$. Nevertheless, the good reference secondary system's performance comes at the cost of a loss of performance at the reference primary system due to the limitation of resources. Such a limitation does not happen to the proposed primary system. This finding becomes even more evident when seen from the sum spectral efficiency point of view in Fig.~\ref{fig:ofdm_vs_vfdm_sum}, where the spectral efficiency of both the primary and secondary systems are summed up. In this result, it becomes evident that the proposed primary-secondary surpasses the performance of the reference primary-secondary, even with the VFDM system under full interference from the primary ($\alpha=1$).

\begin{figure}[!h]
	\centering
	\includegraphics[width=\columnwidth]{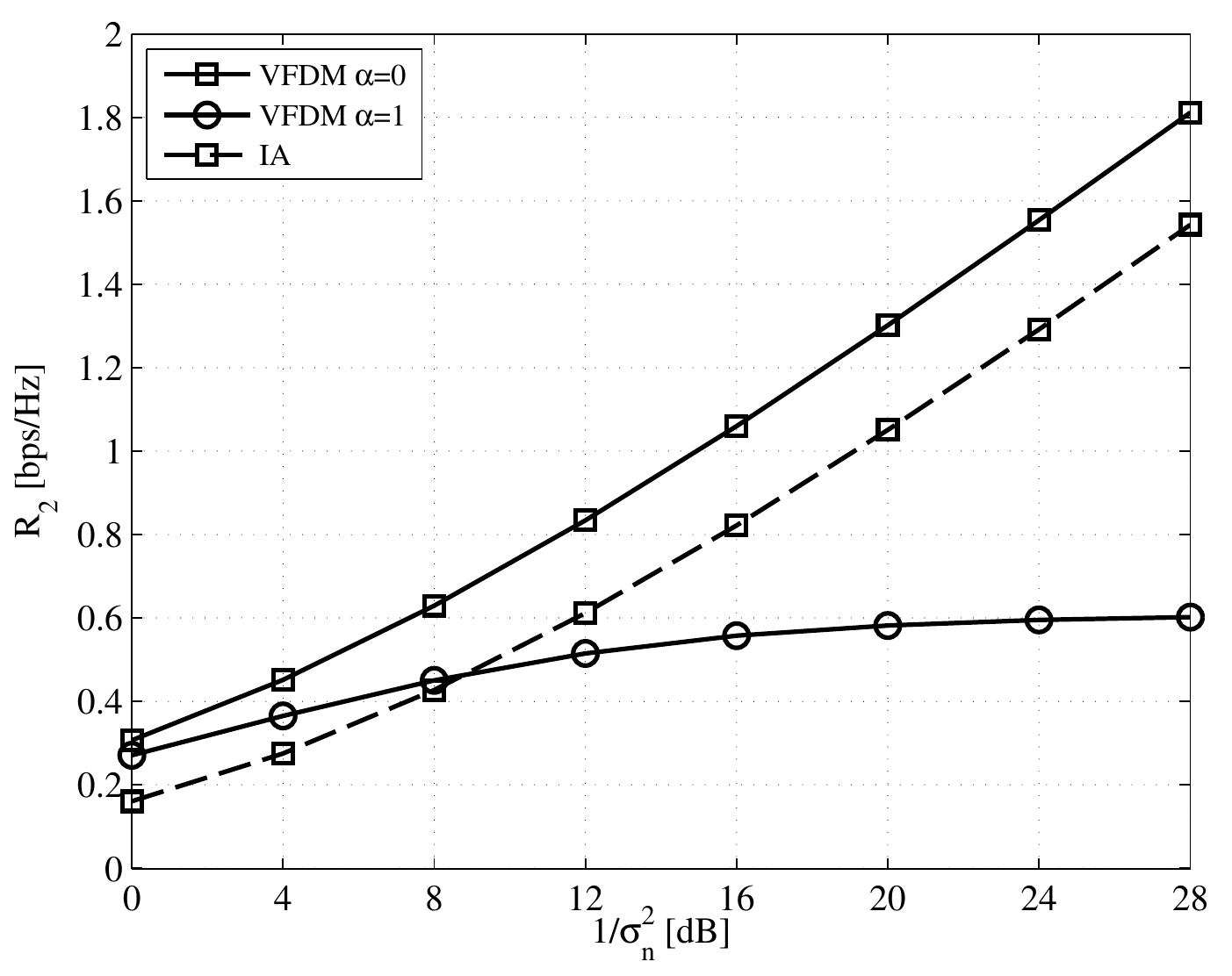} 
	\caption{VFDM's spectral efficiency for $N=64$, $L=16$ and $\alpha \in \{0,1\}$, compared to IA (with equivalent available resources).}
	\label{fig:ia_vs_vfdm}
\end{figure}

We have also compared VFDM to an OFDM based IA~\cite{inpr:shen2008}, in spite of the fact that such an IA proposal is not fit for the kind of scenario targeted by this work. We remark that this comparison is made only to benchmark VFDM as a \emph{mathematical tool}, rather than a practical technique for cognitive interference channels. Before we start, some remarks are of order. We have decided not to include the primary system performances since, it is known for the VFDM case, and it is the same for the primary and secondary receivers in the OFDM based IA. Indeed, IA is a symmetric technique, in the sense that it cancels the interference to both receivers simultaneously. We have considered the OFDM based IA to use  an equivalent amount of resources (rank of the IA precoder is $L$) for fairness. The rest of the parameters are the same for both systems. In Fig.~\ref{fig:ia_vs_vfdm} we see the rate at the secondary system only for both VFDM and the OFDM based IA. The OFDM based IA clearly outperforms VFDM for the full interference case ($\alpha=1$) for SNRs above 9~dB. Since the OFDM based IA exploits cooperation to cancel the interference to both receivers, it is not interference limited as VFDM. Nevertheless, for SNRs below 9~dB, VFDM outperforms the OFDM based IA due to its more robust symbol protection, afforded by the redundancy of the transmitted block. When VFDM is not interference limited (for $\alpha=0$) it outperforms the OFDM based IA throughout the whole SNR range with a constant gain of about 0.2~bps/Hz, due to the losses imposed by the zero-forcing decoder at the IA's receiver. However this kind of situation is unlikely to happen in a realistic scenario.

\subsection{Imperfect Channel Estimation}

One key parameter for implementing wireless systems is the correct proportion between training and data symbols. In this part we present our findings based on the channel estimation protocol described in Sec.~\ref{sec:channel_est}. Particularly for these results, a coherence time of $T=6400$ symbols is considered. As specified in Sec.~\ref{sec:channel_est}, the minimum training size is of $\tau = N$ and the maximum is $\tau = T-N$, even though we concentrate on the initial region of the curves in order to emphasize the best ratio between training and transmit symbols $\tau/T$ region. The performance of channel estimation in terms or spectral efficiency is dependent on two main parts: a) the pre-log factor, a multiplicative linear factor that depends directly on the amount of training symbols $\tau$ and b) the in-log factor, varying with respect to the SNR given by the channel estimate, which is also a function of $\tau$.

\begin{figure}[!htb]
	\centering
	\includegraphics[width=\columnwidth]{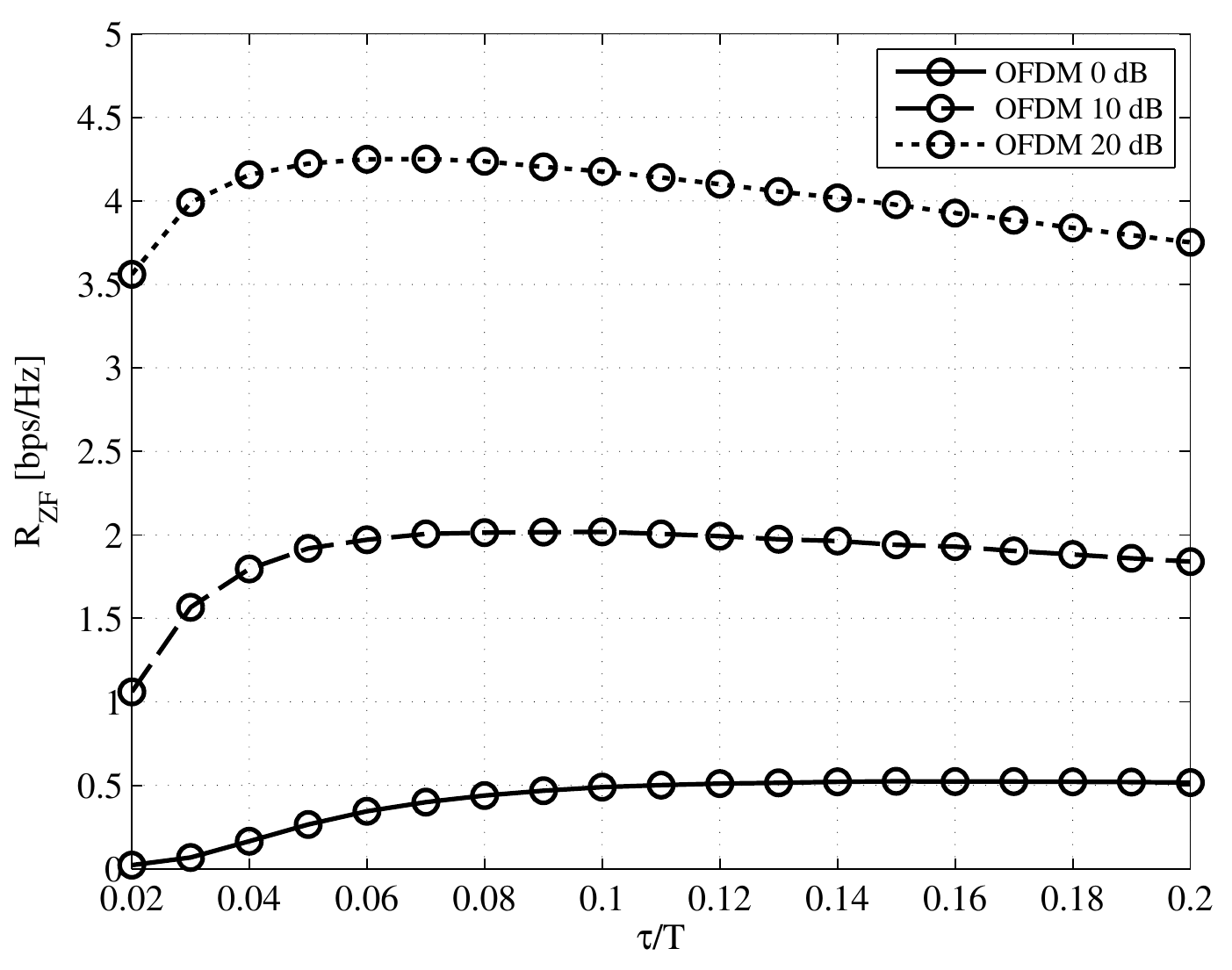} 
   	\caption{Comparative probability of bit error ($P_{e}$) for the MMSE and ZF equalizer for the secondary (VFDM) system with varying interference levels $\alpha = \{0, 0.5, 1\}$ ($N=64$, $L=16$).}
	\label{fig:ofdm_imperf}
\end{figure}

\begin{figure}[!htb]
	\centering
	\includegraphics[width=\columnwidth]{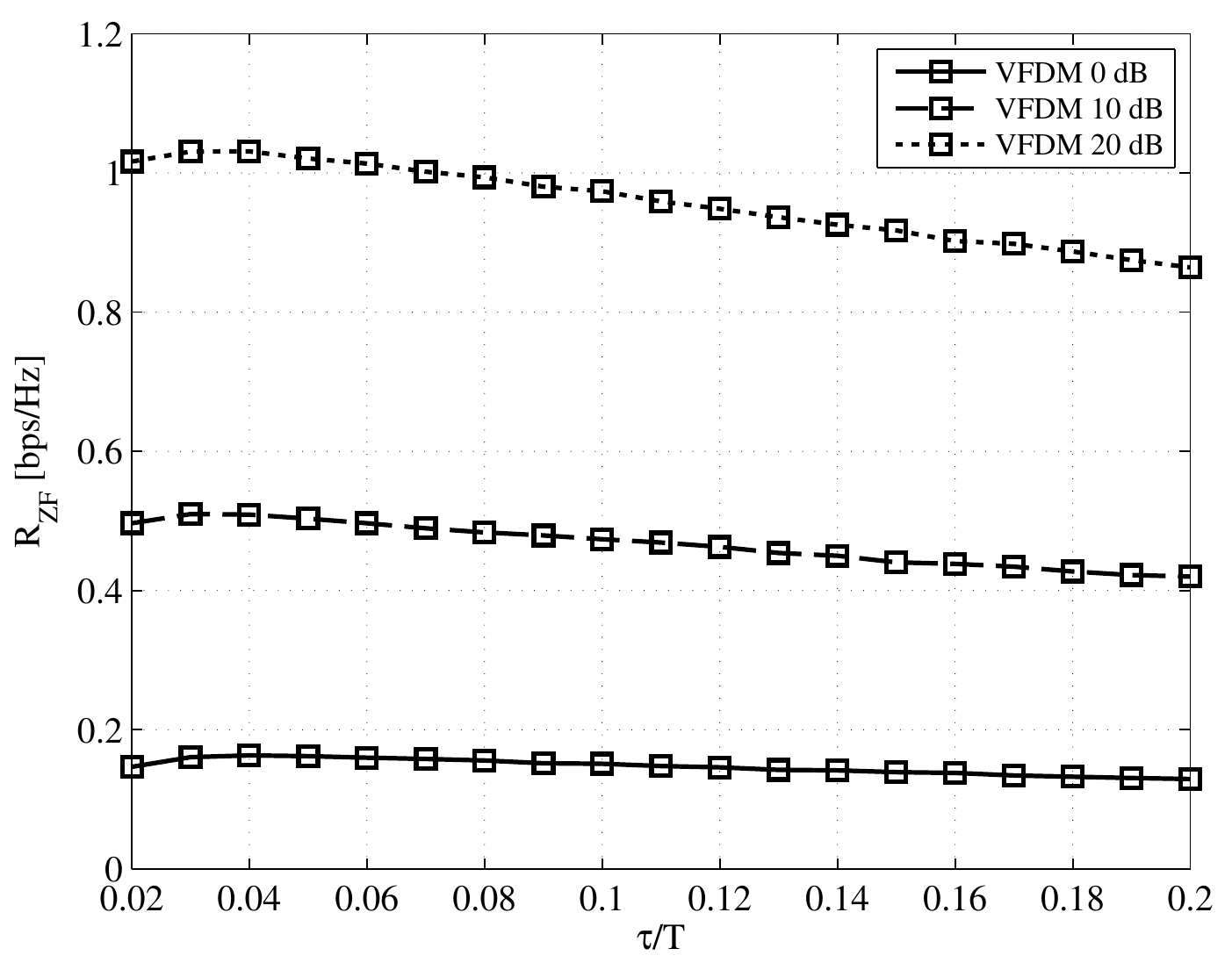} 
   	\caption{Comparative probability of bit error ($P_{e}$) for the MMSE and ZF equalizer for the secondary (VFDM) system with varying interference levels $\alpha = \{0, 0.5, 1\}$ ($N=64$, $L=16$).}
	\label{fig:vfdm_imperf}
\end{figure}

In Fig.~\ref{fig:ofdm_imperf}, the impact of imperfect channel estimation, as a function of $\tau/T$ for the primary system, is presented. The initial part of the curves, before the maximum point (better seen in the 10 and 20~dB curves), is affected mainly by the in-log effect of the imperfect channel estimation on the SNR. From the maximum and on, the pre-log factor kicks in and the reduction in rate is predominantly linear, due to the exchange in the amount of data symbols in favor of more training. As expected, the lower the SNR, the more training symbols are needed to provide a better estimation. As the SNR increases, less symbols are needed. The same behavior can be seen for the secondary system, in Fig.~\ref{fig:vfdm_imperf}. Unlike the primary system, the secondary system's performance is not dependent on the SNR and achieves the best spectral efficiency at a very low $\tau/T$.

\section{Conclusions}\label{sec:conclusions}

In this work, an overlay technique that exploits the frequency selectivity of channels to achieve spectrum sharing, called VFDM, has been introduced. VFDM creates a precoder orthogonal to the interfering channel that achieves interference mitigation. We have shown how such a precoder can be constructed and analyzed its achievable rate performance. A practical channel estimation procedure is introduced, and the best proportion of training versus transmission symbols is analyzed. Throughout this work, the use of numerical examples help to show that, even though VFDM's performance is constrained by the size of the Vandermonde-subspace (null-space) of the interfering channel between the secondary transmitter and primary receiver, non-negligible rates can be achieved.

The extension of VFDM to the multi-user scenario, as well as the implementation of a VFDM testbed is currently under submission. A solution for the general channel distribution problem is currently under studies, and will be a subject of a future publication. 

As a future perspective, we plan to take on the problem of a VFDM system level deployment strategy, dealing with issues such as inter-cell interference management and uplink, among others. A study to determine what level of interconnection is necessary to achieve a target performance will be the focus of a cooperation-to-performance tradeoff study. Finally, techniques able to deal with the limited backhaul secondary system case will be studied.

\balance

\bibliographystyle{IEEEtran}
\bibliography{vfdm}

\end{document}